\documentclass[12pt]{article}

\usepackage{amsfonts}
\usepackage{amssymb}
\usepackage{amscd}

\def\D{\hbox{D\kern-.73em\raise.25ex\hbox{-}\raise-.25ex\hbox{ }}}
 \def\d{\hbox{d\kern-.33em\raise.75ex\hbox{-}\raise-.75ex\hbox{}}}

\def\GGG{\frak G }
\def\gr3{\GGG\,(\SSS_3)}

\def\gr2{\GGG\,(\SSS_2)}

\def\SSS{\frak S}

\def\ed{\end{document}}
\def\beq{\begin{equation}}
\def\eeq{\end{equation}}
\def\bea{\begin{eqnarray}}
\def\eea{\end{eqnarray}}
\def\ba{\begin{array}}
\def\ea{\end{array}}
\def\bi{\begin{itemize}}
\def\ei{\end{itemize}}

\newcommand{\bp}{\noindent\begin{minipage}[c]}
\newcommand{\ep}{\end{minipage}}

\begin{document}
\title{\bf On $p$-Adic Sector of Adelic String}


\author{Branko Dragovich\thanks{\,e-mail
address: dragovich@phy.bg.ac.yu} \\ {}\\
\it{Institute of Physics}\\ \it{Pregrevica 118, P.O. Box 57, 11001
Belgrade, Serbia}}

\date {~}
\maketitle
\begin{abstract}
We consider construction of Lagrangians which are candidates for
$p$-adic sector of an adelic open scalar string. Such Lagrangians
have their origin in Lagrangian for a single $p$-adic string and
contain the Riemann zeta function with the d'Alembertian in its
argument. In particular, we present a new Lagrangian obtained by
an additive approach which takes into account all $p$-adic
Lagrangians. The very attractive feature of this new Lagrangian is
that it is an analytic function of the d'Alembertian.
Investigation of the field theory with Riemann zeta function is
interesting in itself as well.
\end{abstract}
\bigskip

\section{Introduction}

The first notion of a $p$-adic string was introduced by I. V.
Volovich in 1987 \cite{volovich1}. After that, various versions of
$p$-adic strings were developed. The most interest have attracted
strings whose only world sheet is $p$-adic and all other
properties are described by real or complex numbers. Such $p$-adic
strings are connected with ordinary ones by  product of their
scattering amplitudes and  notion of an adelic string has been
considered. Adelic string enables to treat ordinary and $p$-adic
strings simultaneously and on an equal footing. Adelic strings can
be regarded as  more fundamental than ordinary and $p$-adic ones
(for a review of the early days developments, see e.g.
\cite{freund1,volovich2}). Some $p$-adic structures have been also
observed in many other parts of modern mathematical physics (for a
recent review we refer to \cite{dragovich}).

One of the greatest achievements in $p$-adic string theory is an
effective field description of open scalar $p$-adic string tachyon
\cite{freund2,frampton1}. The corresponding Lagrangian is
nonlocal, nonlinear, simple and exact. It describes four-point
scattering amplitudes as well as all higher ones at the
tree-level.

In the last decade the Lagrangian approach to $p$-adic string
theory has been significantly advanced and many aspects of
$p$-adic string dynamics have been investigated, compared with
dynamics of ordinary strings and applied to nonlocal cosmology
(see, e.g. \cite{sen,zwiebach,vladimirov,arefeva,barnaby} and
references therein).

Adelic approach to the string scattering amplitudes connects
$p$-adic and ordinary counterparts, eliminates unwanted prime
number parameter $p$ contained in $p$-adic amplitudes and  cures
the problem of $p$-adic causality violation. Adelic quantum
mechanics \cite{dragovich2} was also successfully formulated, and
it was found a connection between adelic vacuum state of the
harmonic oscillator and the Riemann zeta function. There is also
successful application of adelic analysis to Feynman path integral
\cite{dragovich-b}, quantum cosmology \cite{dragovich-c},
summation of divergent series \cite{dragovich-d}, and dynamical
systems \cite{dragovich-e}.

The present paper is a result of investigation towards
construction of an effective field theory Lagrangian  for $p$-adic
sector of adelic  open scalar string. At the beginning, we give a
brief review of Lagrangian for $p$-adic string and also of our
previous work on this subject. Then, we present a new Lagrangian,
which also contains Riemann zeta function, but in such way that
Lagrangian is now an analytic function of the d'Alembertian
$\Box$.
 Note that $p$-adic sector  of the four point adelic string amplitude contains
the Riemann zeta function.

\section{$p$-Adic and Adelic Strings}

Let us recall the crossing symmetric Veneziano amplitude for
scattering of two ordinary open strings:
\begin{equation}
A_\infty (a, b) = g_\infty^2\, \int_{\mathbb{R}}
|x|_{\infty}^{a-1} \, |1-x|_{\infty}^{b-1}\, d_\infty x =
g_\infty^2 \frac{\zeta (1-a)}{\zeta (a)}\, \frac{\zeta
(1-b)}{\zeta (b)}\, \frac{\zeta (1-c)}{\zeta (c)}\,, \label{1.1}
\end{equation}
where $a = - \alpha (s) = - \frac{s}{2} - 1,\, b = - \alpha (t),\,
c = - \alpha (u)$ with the condition $a + b + c = 1$, i.e. $s + t
+ u = - 8$. In (\ref{1.1}), $\, |\cdot|_\infty$ denotes the
ordinary absolute value, $\mathbb{R}$ is the field of real
numbers, kinematic variables $a, b, c \in \mathbb{C}$, and $\zeta$
is the Riemann zeta function. The corresponding Veneziano
amplitude for scattering of $p$-adic strings was introduced  as
$p$-adic analog of the integral in (\ref{1.1}), i.e.
\begin{equation}
A_p (a, b) = g_p^2\, \int_{\mathbb{Q}_p} |x|_p^{a-1} \,
|1-x|_p^{b-1}\, d_p x \,, \label{1.2}
\end{equation}
where $\mathbb{Q}_p$ is the field of $p$-adic numbers, $|\cdot
|_p$ is $p$-adic absolute value and $d_p x$ is the additive Haar
measure on $\mathbb{Q}_p$. In (\ref{1.2}), kinematic variables $a,
b, c$ maintain their complex values with condition $a + b + c =
1$. After integration in (\ref{1.2}) one obtains
\begin{equation}
A_p (a, b) = g_p^2\, \frac{1- p^{a-1}}{1- p^{-a}}\, \frac{1-
p^{b-1}}{1- p^{-b}}\,\frac{1- p^{c-1}}{1- p^{-c}}\,, \label{1.3}
\end{equation}
where $p$ is any prime number. Recall the definition of the
Riemann zeta function
\begin{equation}
\zeta (s) = \sum_{n= 1}^{+\infty} \frac{1}{n^{s}} = \prod_p
\frac{1}{ 1 - p^{- s}}\,, \quad s = \sigma + i \tau \,, \quad
\sigma
>1\,, \label{1.4}
\end{equation}
which has analytic continuation to the entire complex $s$ plane,
excluding the point $s=1$, where it has a simple pole with residue
1.  According to (\ref{1.4}) one can take product of $p$-adic
string amplitudes
\begin{equation}
\prod_p A_p (a, b) =  \frac{\zeta (a)}{\zeta (1-a)}\, \frac{\zeta
(b)}{\zeta (1-b)}\, \frac{\zeta (c)}{\zeta (1-c)} \, \prod_p
g_p^2\,, \label{1.5}
\end{equation}
what gives a nice simple  formula
\begin{equation}
A_\infty (a, b) \, \prod_p A_p (a, b) = g_\infty^2 \, \prod_p
g_p^2\,. \label{1.6}
\end{equation}
To have infinite product of amplitudes (\ref{1.6}) finite it must
be finite product of coupling constants, i.e. $g_\infty^2 \,
\prod_p g_p^2 = const.$ From (\ref{1.6}) it follows that the
ordinary Veneziano amplitude, which is rather complex, can be
expressed as product of all inverse $p$-adic counterparts, which
are much more simpler. Moreover, expression (\ref{1.6}) gives rise
to consider it as the amplitude for an adelic string, which is
composed of the ordinary and $p$-adic ones.

\subsection{Lagrangian for a $p$-Adic Open String}

The exact tree-level Lagrangian of the effective scalar field
$\varphi$, which describes the open $p$-adic string tachyon, is
\cite{freund2,frampton1}
\begin{equation} {\cal L}_p = \frac{m^D}{g_p^2}\, \frac{p^2}{p-1} \Big[
-\frac{1}{2}\, \varphi \, p^{-\frac{\Box}{2 m^2}} \, \varphi  +
\frac{1}{p+1}\, \varphi^{p+1} \Big]\,,  \label{2.1} \end{equation}
where $p$
 is a prime, $\Box = - \partial_t^2  + \nabla^2$ is the
$D$-dimensional d'Alembertian.

An infinite number of spacetime derivatives follows from the
expansion
$$
p^{-\frac{\Box}{2 m^2}} = \exp{\Big( - \frac{1}{2 m^2} \log{p}\,
\Box \Big)} = \sum_{k = 0}^{+\infty} \, \Big(-\frac{\log p}{2 m^2}
\Big)^k \, \frac{1}{k !}\, \Box^k \,.
$$
The equation of motion for (\ref{2.1}) is

\begin{equation} p^{-\frac{\Box}{2 m^2}}\, \varphi = \varphi^p \,,
\label{2.4} \end{equation} and its properties have been studied by
many authors (see, \cite{vladimirov} and references therein).

\section{Lagrangians for $p$-Adic Sector}

Now  we want to consider construction of Lagrangians which  are
candidates to describe entire $p$-adic sector of an adelic open
scalar string. In particular, an appropriate such Lagrangian
should describe scattering amplitude (\ref{1.5}), which contains
the Riemann zeta function. Consequently, this Lagrangian  has to
contain the Riemann zeta function with the d'Alembertian in its
argument. Thus we have to look for possible constructions of a
Lagrangian which contains the Riemann zeta function and has its
origin in $p$-adic Lagrangian (\ref{2.1}). We have found and
considered two approaches: additive and multiplicative.

\subsection{Additive approach}

Prime number $p$ in (\ref{2.1})  can be replaced by any natural
number $n \geq 2$ and consequences  also make sense.

Now we want to introduce a Lagrangian which incorporates all the
above  Lagrangians (\ref{2.1}), with $p$ replaced by $n \in
\mathbb{N}$. To this end, we take the sum of all Lagrangians
${\cal L}_n$  in the form

\begin{equation} L =   \sum_{n = 1}^{+\infty} C_n\, {\cal L}_n   =
 \sum_{n= 1}^{+\infty} C_n \frac{ m^D}{g_n^2}\frac{n^2}{n -1}
\Big[ -\frac{1}{2}\, \phi \, n^{-\frac{\Box}{2 m^2}} \, \phi +
\frac{1}{n + 1} \, \phi^{n+1} \Big]\,, \label{3.1}
\end{equation} whose explicit realization depends on particular
choice of coefficients $C_n$ and coupling constants $g_n$. To
avoid a divergence  in $1/(n-1)$ when $n = 1$ one has to take that
${C_n}/{g_n^2}$ is proportional to $n -1$. Here we consider some
cases when coefficients $C_n$ are proportional to $n-1$, while
coupling constants $g_n$ do not depend on $n$, i.e. $  g_n = g$.
In fact, according to (\ref{1.6}), in this case $  g_n^2 = g^2 =
1$. Another possibility is that $C_n$ is not proportional to
$n-1$, but $g_n^2 = \frac{n^2}{n^2 - 1}$ and then $\prod_p g_p^2 =
\zeta (2) = \frac{\pi^2}{6}$, what is consistent with (\ref{1.6}).
To differ this new field from a particular $p$-adic one, we use
notation $\phi$ instead of $\varphi$.

We have considered three cases for coefficients $C_n$ in
(\ref{3.1}): (i) $C_n = \frac{n-1}{n^{2+h}}$, where $h$ is a real
parameter; (ii) $C_n = \frac{n^2 -1}{n^2}$; and (iii) $C_n = \mu
(n)\, \frac{n-1}{n^2}$, where $\mu (n)$ is the M\"obius function.

Case (i) was considered in \cite{dragovich3,dragovich4}. Obtained
Lagrangian is
\begin{equation} L_{h} =  \frac{m^D}{g^2} \Big[ \,- \frac{1}{2}\,
 \phi \,  \zeta\Big({\frac{\Box}{2 \, m^2}  +
h }\Big) \, \phi    + {\cal{AC}} \sum_{n= 1}^{+\infty} \frac{n^{-
h}}{n + 1} \, \phi^{n+1} \Big]\,, \label{3.2} \end{equation} where
$\mathcal{AC}$ denotes analytic continuation.

Case (ii) was investigated in \cite{dragovich5} and the
corresponding Lagrangian is

 \begin{equation} L =  \frac{m^D}{g^2} \Big[ \, - \frac{1}{2}\,
 \phi \,  \Big\{ \zeta\Big({\frac{\Box}{2\, m^2}  -
 1}\Big)\, + \, \zeta\Big({\frac{\Box}{2\, m^2} }\Big) \Big\} \, \phi \,  + \,   \frac{\phi^2}{1 - \phi} \,
 \Big]\,. \label{3.3} \end{equation}

Case with the M\"obius function $\mu (n)$ is presented in
\cite{dragovich6} and  the corresponding Lagrangian is

\begin{equation}
L =   \frac{m^D}{g^2} \Big[ - \frac{1}{2}\, \phi \, \frac{1}{
\zeta\Big({\frac{\Box}{2 m^2}}\Big)} \,\phi + \int_0^\phi {\cal
M}(\phi) \, d\phi\Big] , \label{3.6}
\end{equation}
where ${\cal M}(\phi) = \sum_{n= 1}^{+\infty} {\mu (n)} \,
\phi^{n} = \phi - \phi^2 - \phi^3 - \phi^5 + \phi^6 - \phi^7 +
\phi^{10} - \phi^{11} - \dots $.

\subsection{Multiplicative approach}

In the multiplicative approach the Riemann zeta function  emerges
through its product form (\ref{1.4}). Our starting point is again
$p$-adic Lagrangian (\ref{2.1}). It is useful to rewrite
(\ref{2.1}), first in the form,
\begin{equation} {\cal L}_p = \frac{m^D}{g_p^2}\, \frac{p^2}{p^2-1} \Big\{
-\frac{1}{2}\, \varphi \, \Big[ p^{-\frac{\Box}{2 m^2}+1} +
p^{-\frac{\Box}{2 m^2}} \Big]\, \varphi + \, \varphi^{p+1}
\Big\}\, \label{3.2.1}
\end{equation}
and then, by addition  and substraction of $\varphi^2$, as
\begin{equation} {\cal L}_p = \frac{m^D}{g_p^2}\, \frac{p^2}{p^2-1} \Big\{
\frac{1}{2}\, \varphi \, \Big[ \Big(1 - p^{-\frac{\Box}{2 m^2}+1}
\Big) + \Big( 1 - p^{-\frac{\Box}{2 m^2}}\Big) \Big]\, \varphi -
\varphi^2 \Big( 1 - \varphi^{p-1} \Big) \Big\}\,. \label{3.2.2}
\end{equation}

Taking products
\begin{equation}
 \prod_p g_p^2  = C \,, \, \prod_p \frac{1}{1 - p^{-2}}\,, \, \prod_p (1 - p^{-\frac{\Box}{2 m^2}+1}) \,,
 \, \prod_p (1 - p^{-\frac{\Box}{2 m^2}}) \,, \, \prod_p ( 1 - \varphi^{p-1})  \label{3.2.3}
\end{equation}
in (\ref{3.2.2}) at the relevant places one obtains Lagrangian
\begin{equation}
{\mathcal L} = \frac{m^D}{C}\, \zeta (2)\, \Big\{ \frac{1}{2} \,
\phi \Big[ \zeta^{-1} \Big( \frac{\Box}{2 m^2} - 1 \Big) +
\zeta^{-1} \Big( \frac{\Box}{2 m^2}  \Big)\Big] \, \phi - \phi^2
\prod_p \Big( 1 - \phi^{p-1} \Big) \Big\} \,,  \label{3.2.4}
\end{equation}
where $\zeta^{-1} (s) = 1/\zeta (s)$. It is worth noting that from
Lagrangian (\ref{3.2.4}) one can easily reproduce its $p$-adic
ingredient (\ref{3.2.1}). Lagrangian (\ref{3.2.4}) was introduced
and considered in \cite{dragovich7}. In particular, it was shown
that very similar Lagrangian can be obtained from the additive
approach with the M\"obius function and that these two Lagrangians
describe the same field theory in the week field approximation.

\subsection{A new Lagrangian with Riemann zeta function}

Here we present a new Lagrangian constructed by additive approach
taking $C_n = (-1)^{n-1}\, \frac{n^2 -1}{n^2}$ in (\ref{3.1}).
This choice of coefficients $C_n$ is similar to the above case
(ii) and  distinction is in the sign $(-1)^{n-1}$. The starting
$p$-adic Lagrangian is in the form (\ref{3.2.1}) and it gives
\begin{equation} L =
 \sum_{n= 1}^{+\infty} C_n \frac{ m^D}{g_n^2}\frac{n^2}{n^2 -1}
\Big[ -\frac{1}{2}\, \phi \, n^{-\frac{\Box}{2 m^2} +1} \, \phi
-\frac{1}{2}\, \phi \, n^{-\frac{\Box}{2 m^2}} \, \phi +  \,
\phi^{n+1} \Big]\,. \label{3.3.1}
\end{equation}

Recall that
\begin{equation}
\sum_{n= 1}^{+\infty} (-1)^{n-1} \frac{1}{n^{s}} =  (1 - 2^{1-s})
\, \zeta (s), \quad s = \sigma + i \tau \,, \quad \sigma
> 0\,, \label{3.3.2}
\end{equation}
which has analytic continuation to the entire complex $s$ plane
without singularities. At point $s = 1$, one has $\lim_{s\to 1} (1
- 2^{1-s}) \, \zeta (s)\, = \, \sum_{n= 1}^{+\infty} (-1)^{n-1}
\frac{1}{n} \, = \, \log 2$. Applying (\ref{3.3.2}) to
(\ref{3.3.1}) and using analytic continuation one obtains
 \begin{equation} L = - {m^D} \Big[ \,  \frac{1}{2}\,
 \phi \,  \Big\{ \, \Big(1 - 2^{2 - \frac{\Box}{2 m^2}}\Big)\, \zeta\Big({\frac{\Box}{2\, m^2}  -
 1}\Big)\, + \,  \Big(1 - 2^{1 - \frac{\Box}{2 m^2}}\Big)\, \zeta\Big({\frac{\Box}{2\, m^2} }\Big)
 \Big\} \, \phi \,  - \,   \frac{\phi^2}{1 + \phi} \,
 \Big]\,, \label{3.3.3} \end{equation}
where it was taken $g_n^2 = g^2 = 1$.

The corresponding equation of motion is
\begin{equation}
   \Big[ \, \Big(1 - 2^{2 - \frac{\Box}{2 m^2}}\Big)\, \zeta\Big({\frac{\Box}{2\, m^2}  -
 1}\Big)\, + \,  \Big(1 - 2^{1 - \frac{\Box}{2 m^2}}\Big)\, \zeta\Big({\frac{\Box}{2\, m^2} }\Big)
 \Big] \, \phi \,   = \,   \frac{\phi^2 + 2 \phi}{(1 + \phi)^2} \,, \label{3.3.4} \end{equation}
which in the week field approximation  gives  equation
\begin{equation}
    \Big(1 - 2^{2 - \frac{M^2}{2 m^2}}\Big)\, \zeta\Big({\frac{M^2}{2\, m^2}  -
 1}\Big)\, + \,  \Big(1 - 2^{1 - \frac{M^2}{2 m^2}}\Big)\, \zeta\Big(\frac{M^2}{2\, m^2} \Big) - 2 = 0
    \, \label{3.3.5} \end{equation}
    for the spectrum of masses $M^2$ as function of string mass $m^2$. Equation (\ref{3.3.4}) has three $\phi=const.$ solutions,
    which are $\phi = 0,\, 1, \, -\frac{5}{3}$.

 The potential can be obtained by equality $V (\phi) = - L (\Box = 0)$, i.e.
 \begin{equation}
V (\phi) = m^D \, \frac{3 \,\phi - 5}{8 \,(1 + \phi)}\,  \phi^2
\label{3.3.6}
 \end{equation}
which has two local minima at $\phi = 1$ and $\phi = -
\frac{5}{3}$, and it has one local maximum $V(0) = 0$. These
values of $\phi$
 coincide with constant solutions  of equation of motion (\ref{3.3.4}).
Potential (\ref{3.3.6}) is singular at $\phi = -1$. Note that sign
$(-1)^{p-1}$ in front of ${\cal L}_p$ in (\ref{3.3.1}) is positive
when $p$ is an odd prime and it has as a result that $V (\phi) \to
+ \infty$ when $\phi \to \pm \infty$.

\section{Concluding remarks}

The main result of this paper is construction of the Lagrangian
(\ref{3.3.3}). Unlike previously constructed Lagrangians, this one
has no singularity with respect to the d'Alembertian $\Box$ and it
enables to apply easier pseudodifferential approach.  This
analyticity of the Lagrangian should be also useful in its
application to nonlocal cosmology, which uses  linearization
procedure (see, e.g. \cite{koshelev} and references therein).

It is worth mentioning that an interesting approach towards
foundation of a field theory and cosmology based on the Riemann
zeta function was proposed in \cite{volovich3}.

\section*{\large Acknowledgements}
The work on this article was partially supported by the Ministry
of Science and Technological Development, Serbia, under contract
No 144032D. The author  thanks organizers of the SFT'09 conference
at the Steklov Mathematical Institute, Moscow, for a stimulating
and useful scientific meeting.

 \end{document}